\documentstyle[epsfig]{mn}
\input{psfig}

\newif\ifAMStwofonts

\ifoldfss
  \ifCUPmtlplainloaded \else
    \NewTextAlphabet{textbfit} {cmbxti10} {}
    \NewTextAlphabet{textbfss} {cmssbx10} {}
    \NewMathAlphabet{mathbfit} {cmbxti10} {} 
    \NewMathAlphabet{mathbfss} {cmssbx10} {} 
  \fi
  \ifAMStwofonts
    \ifCUPmtlplainloaded \else
      \NewSymbolFont{upmath} {eurm10}
      \NewSymbolFont{AMSa} {msam10}
      \NewMathSymbol{\upi}     {0}{upmath}{19}
      \NewMathSymbol{\umu}     {0}{upmath}{16}
      \NewMathSymbol{\upartial}{0}{upmath}{40}
      \NewMathSymbol{\leqslant}{3}{AMSa}{36}
      \NewMathSymbol{\geqslant}{3}{AMSa}{3E}

       \let\le=\leqslant
      \let\geq=\geqslant \let\ge=\geqslant
    \fi
  \fi
\fi 

\ifnfssone
  \newmathalphabet{\mathit}
  \addtoversion{normal}{\mathit}{cmr}{m}{it}
  \addtoversion{bold}{\mathit}{cmr}{bx}{it}
  \newmathalphabet{\mathbfit} 
  \addtoversion{normal}{\mathbfit}{cmr}{bx}{it}
  \addtoversion{bold}{\mathbfit}{cmr}{bx}{it}
  \newmathalphabet{\mathbfss} 
  \addtoversion{normal}{\mathbfss}{cmss}{bx}{n}
  \addtoversion{bold}{\mathbfss}{cmss}{bx}{n}
  \ifAMStwofonts
    \ifCUPmtlplainloaded \else
      \UseAMStwoboldmath
      \makeatletter
      \new@mathgroup\upmath@group
      \define@mathgroup\mv@normal\upmath@group{eur}{m}{n}
      \define@mathgroup\mv@bold\upmath@group{eur}{b}{n}
      \edef\UPM{\hexnumber\upmath@group}
      \new@mathgroup\amsa@group
      \define@mathgroup\mv@normal\amsa@group{msa}{m}{n}
      \define@mathgroup\mv@bold\amsa@group{msa}{m}{n}
      \edef\AMSa{\hexnumber\amsa@group}
      \makeatother
      \mathchardef\upi="0\UPM19
      \mathchardef\umu="0\UPM16
      \mathchardef\upartial="0\UPM40
      \mathchardef\leqslant="3\AMSa36
      \mathchardef\geqslant="3\AMSa3E

       \let\le=\leqslant
      \let\geq=\geqslant \let\ge=\geqslant
    \fi
  \fi
\fi 

\ifnfsstwo
  \DeclareMathAlphabet{\mathbfit}{OT1}{cmr}{bx}{it}
  \SetMathAlphabet\mathbfit{bold}{OT1}{cmr}{bx}{it}
  \DeclareMathAlphabet{\mathbfss}{OT1}{cmss}{bx}{n}
  \SetMathAlphabet\mathbfss{bold}{OT1}{cmss}{bx}{n}
  \ifAMStwofonts
    \ifCUPmtlplainloaded \else
      \DeclareSymbolFont{UPM}{U}{eur}{m}{n}
      \SetSymbolFont{UPM}{bold}{U}{eur}{b}{n}
      \DeclareSymbolFont{AMSa}{U}{msa}{m}{n}
      \DeclareMathSymbol{\upi}{0}{UPM}{"19}
      \DeclareMathSymbol{\umu}{0}{UPM}{"16}
      \DeclareMathSymbol{\upartial}{0}{UPM}{"40}
      \DeclareMathSymbol{\leqslant}{3}{AMSa}{"36}
      \DeclareMathSymbol{\geqslant}{3}{AMSa}{"3E}

       \let\le=\leqslant
      \let\geq=\geqslant \let\ge=\geqslant
    \fi
  \fi
\fi 

\ifCUPmtlplainloaded \else
  \ifAMStwofonts \else 
    \def\upi{\pi}
    \def\umu{\mu}
    \def\upartial{\partial}
  \fi
\fi

\newcommand{\einstein}{{\it Einstein} }
\newcommand{\rosat}{{\it ROSAT} }
\newcommand{\sax}{{\it BeppoSAX} }

\newcommand{\fxfr}{{$f_{\rm x}/f_{\rm r}$~}}

\newcommand{\ergs}{{erg~cm$^{-2}$~s$^{-1}$}}
\newcommand{\ergj}{{erg~cm$^{-2}$~s$^{-1}$~Jy$^{-1}$}}

\newcommand{\aro} {{$\alpha_{\rm ro}$}}
\newcommand{\aoxaro}{{$\alpha_{\rm ox}-\alpha_{\rm ro}$}}
\newcommand{\lsim}{{\lower.5ex\hbox{$\; \buildrel < \over \sim 
\;$}}}
\newcommand{\gsim}{{\lower.5ex\hbox{$\; \buildrel > \over \sim 
\;$}}}
\newcommand{\vovm}{{$V/V_{\rm m}$}~}
\newcommand{\vova}{{$V_{\rm e}/V_{\rm a}$}~}

\newcommand{\vovmave}{{$\langle V/V_{\rm m} \rangle$}~}
\newcommand{\vovaave}{{$\langle V_{\rm e}/V_{\rm a} \rangle$}~}

\title[The Sedentary Multi-Frequency Survey]
{The Sedentary Multi-Frequency Survey. I. \\
Statistical Identification and Cosmological Properties of  
HBL BL Lacs}
\author[P. Giommi, M.T. Menna, P. Padovani]
       {P. Giommi$^{1,2}$, M.T. Menna$^1$, P. Padovani$^{3,4,5}$ \\
       $^1$ \sax Science Data Center, ASI, 
    Via Corcolle, 19
    I-00131 Roma, Italy \\
      $^2$ Area per la Ricerca Scientifica, Italian Space Agency \\
      $^3$ Space Telescope Science Institute,
    3700 San Martin Drive,
    Baltimore,  MD 21218
    U.S.A\\
$^4$ Affiliated to the Astrophysics Division, Space Science Department, 
European Space Agency \\
      $^5$ On leave of absence from Dipartimento di Fisica, II 
Universit\`a di Roma ``Tor Vergata'', Italy
}
\date{
      Received 1999 ;
      in original form 1999 }

\pagerange{\pageref{firstpage}--\pageref{lastpage}}
\pubyear{1999}

\begin{document}
\maketitle

\label{firstpage}

\begin{abstract}

We have assembled a multi-frequency database by cross-correlating the
NVSS catalog of radio sources with the RASSBSC list of soft X-ray
sources, obtaining optical magnitude estimates from the Palomar and UK
Schmidt surveys as provided by the APM and COSMOS on-line services.
By exploiting the nearly unique broad-band properties of High-Energy
Peaked (HBL) BL Lacs we have statistically identified a sample of 218
objects that is expected to include about $85\%$ of BL Lacs and
that is therefore several times larger than all other published
samples of HBLs.  Using a subset (155 objects) that is radio flux
limited and statistically well-defined we have derived the \vovm
distribution and the LogN-LogS of extreme HBLs (\fxfr $\ge 3 \times
10^{-10}$ \ergj) down to 3.5 mJy.  We find that the LogN-LogS flattens
around 20 mJy and that \vovmave $ = 0.42 \pm 0.02$. This extends to the
radio band earlier results, based on much smaller X-ray selected
samples, about the anomalous cosmological observational properties of
HBL BL Lacs.  A comparison with the expected radio LogN-LogS of all BL
Lacs (based on a beaming model) shows that extreme HBLs make up
roughly $2\%$ of the BL Lac population, independently of radio
flux. This result, together with the flatness of the radio logN-logS
at low fluxes, is in contrast with the predictions of a recent model
which assumes an anti-correlation between peak frequency and
bolometric luminosity. This scenario would in fact result in an
increasing dominance of HBLs at lower radio fluxes; an effect that,
if at all present, must start at fluxes fainter than our survey limit. 
The extreme \fxfr
flux ratios and high X-ray fluxes of these BL Lacs makes them good candidate 
TeV sources,
some of the brighter (and closer) ones possibly detectable with the
current generation of Cerenkov telescopes.
Statistical identification of sources based on their location in
multi-parameter space, of the kind described here, will have to become
commonplace with the advent of the many large, deep surveys at various 
frequencies currently scheduled or under construction. 
\end{abstract}


\section{Introduction}

The recent widespread availability of large catalogs of astronomical objects 
(e.g., NVSS, Condon et al. 1998, FIRST, White et al. 1997 in the radio 
band, WGA, White, Giommi \& Angelini 1994, ROSATSRC, 
Voges et al. 1994, and RASSBSC, Voges et al. 1996)
combined with on-line services offering simple access 
to finding charts and magnitude estimates, is providing an 
unprecedented opportunity to carry out large and complex multi-frequency 
surveys, such as, for example, the DXRBS (Perlman et al. 1998) and the REX 
(Caccianiga et al. 1999). 
However, as in all traditional surveys, the initial 
multi-wavelength selection must be followed by optical spectroscopic 
identification of a very large number of candidates. This classical 
approach often requires extremely large amounts of telescope time 
and, consequently, very long times before the survey results can be secured.
An alternative, much faster and by far less demanding, approach is 
the ``statistical identification'' of sources based on 
unique (or nearly unique) characteristics of some classes of sources 
in the multi-dimensional parameter space.
We present here a multi-frequency survey based on archival/catalog 
data that makes use of statistical identification of a type of BL 
Lacertae objects that are known to possess a peculiar broad-band 
energy distribution. 

Since their discovery, BL Lacs stood out from other types of 
extragalactic sources for their extreme and sometimes unique 
properties. These include large and rapid variability,
high polarization, lack of emission lines etc. (e.g., Kollgaard  1994). 
Traditionally, BL Lacs were mostly found in radio surveys, but in the 
1980s the discovery of many BL Lacs as serendipitous sources in 
X-ray images led to the division of this class into Radio selected and 
X-ray selected BL Lacs, the latter apparently showing somewhat 
less extreme properties (e.g., Stocke et al. 1985).  
Very recent observations, however, revealed that 
X-ray selected objects do show very large variability in 
their bolometric luminosity, sometimes accompanied with spectacular spectral 
changes (MKN 501, Pian et al. 1998, 1ES 2344+514, Giommi et al. 1999, MKN 421, 
Malizia et al. 1998, ON 321, Tagliaferri et al. 1998, PKS 2005-489, 
Tagliaferri et al. 1998).

A recent unified model (Padovani \& Giommi 1995b) put forward a 
view where radio and X-ray discovered BL Lacs belong to the same basic 
population and are only distinguished by the peak of their synchrotron 
emission. Objects with synchrotron peak at low energy 
(typically in the Infra-Red) are generally found in radio surveys and are 
called LBLs (low energy peaked BL Lacs), while the rarer objects with 
synchrotron peak in the UV/X-ray band are called HBLs (high 
energy peaked BL Lacs), and are mainly selected in the X-ray band, where
the maximum of their synchrotron power is emitted.

The space density of BL Lacs is intrinsically low. This makes them rare 
objects at nearly all frequencies, except 
in the gamma rays where, together with emission line Blazars, they are by far the most abundant component within the high galactic
latitude population of sources. At even higher energies BL Lacs are the only extragalactic objects to populate the presently known TeV sky.

Synchrotron emission followed by inverse Compton scattering, 
combined with relativistic beaming is generally thought to be the 
mechanism responsible for the smooth emission over such a huge spectral range and for the extreme properties of these singular 
sources (Kollgaard 1994, Urry \& Padovani 1995). 

To date only about 400 BL Lac objects are known (Padovani \& Giommi 1995a, 
Bade et al. 1998, Perlman et al. 1998, Laurent-Muehleisen et al. 1998).
Because of their low surface density, sizable samples of BL Lacs can be 
constructed only from surveys covering large portions of the sky. 

In this paper we report on the first results of a multi-frequency 
search entirely based on radio, optical, and X-ray archival data.
Since this survey does not require optical identification before significant 
conclusions can be drawn, we have called it ``sedentary.''   
In particular, we extract a well defined (radio) flux-limited sample 
of HBLs from which we derive some of the cosmological properties 
of this type of objects. 
The work presented here is complementary to the DXRBS (Perlman et 
al. 1998) which instead concentrates on BL Lacs and other radio loud 
AGN with less extreme values of the X-ray to radio flux ratio (\fxfr) 
and that mainly cover the region of the $\alpha_{\rm ox}-\alpha_{\rm ro}$  
plane occupied by less X-ray bright BL Lacs and by flat-spectrum 
radio quasars (FSRQ).

Throughout this paper we assume a Friedmann cosmology with 
$H_0 =50~Km~s^{-1}~Mpc^{-1}$ and $q_0 = 0$. 
 
\section{The Radio/X-ray cross-correlation}

The NVSS catalog of radio sources (Condon et al. 1998, version 
available 
in June 1998, including 1,807,316 entries), which reaches a minimum flux 
of 2.5 mJy but is complete down to 3.4 mJy, has been cross-correlated
with the RASSBSC catalog (18,811 entries above 0.05 cts/s, Voges et al. 1996) 
of bright 
soft X-ray sources detected during the \rosat All Sky 
Survey.
The correlation has been carried out using the EXOSAT/HEASARC BROWSE 
software as implemented at the \sax Science Data Center (SDC, 
Giommi \& Fiore 1998).
The radius used for the correlation was 0.8 arcminutes, roughly 
corresponding to the largest X-ray position uncertainty for any source 
in the RASSBSC, although 90\% of the error radii are less than 20  
arcseconds in the RASSBSC and typically less than 5 arcseconds in the 
NVSS survey. 

The cross-correlation resulted in 3505 RASSBSC sources 
(2900 of which at $|b|>20^{\circ}$) matching to 3702
entries in the NVSS. In $\sim 5\%$ of the cases (189 sources) more than one
radio source was found within the correlation radius; in these cases we have
assumed that the radio counterpart is the brightest candidate, or the closest
to the X-ray position in case the radio fluxes did not differ by more than a
factor of two. We note that the double-lobed sources discussed by Caccianiga et
al. (1999) would in most cases be missed by our selection procedure. However,
since we are concentrating on BL Lacs, this actually works in the direction
of decreasing the contamination of our sample from objects other than BL Lacs.

Given that the number of sources involved is very large and that 
the X-ray error regions are not always small, we must evaluate the effects of 
spurious spatial coincidences that may occur with non-negligible frequency.
The fraction of spurious matches can be estimated by shifting
the coordinates of all the sources of one of the catalogs by a 
fixed amount and re-run the cross-correlation with the same 
correlation parameters. 
This procedure has been applied four times with different amounts 
of the coordinates shift. 
The results are reported in Table 1 where we see that the expected 
percentage of spurious matches in our fixed radius cross-correlation 
is around $10-13\%$. 

\begin{table}
\begin{center}
\begin{tabular}{|c|c|c|}
\multicolumn{3}{l}{Table~1: Spurious matches in the NVSS/RASSBSC}\\
\multicolumn{3}{l}{cross-correlation}\\
\hline 
Coordinates shift & no. of spurious & percentage\\
degrees &matches &\\ \\[-3mm]
\hline \\[-3mm]
 ~2.0 & 470 & 12.7 \\
 -2.0 & 380 &10.3  \\
 ~4.0 & 460 & 12.4\\
 -4.0 & 421 & 11.4\\
\hline
\end{tabular} 
\end{center}
\end{table}

\begin{figure}
\epsfig{figure=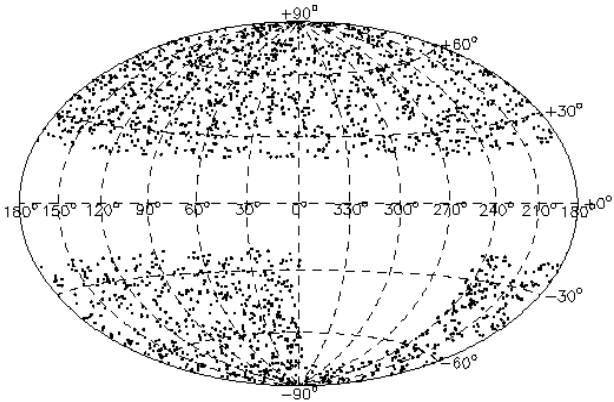,height=5.5cm,width=8.5cm}
\caption{The 2074 NVSS/RASSBSC high-latitude ($|b| > 20^{\circ}$), 
point-like (RASSBSC extent $<$ 30 arcseconds) sources for which an estimate
of the optical magnitude is available are plotted in galactic 
coordinates. The portion of sky covered reflects the 
coverage of the NVSS ($\delta > -40^{\circ}$) and of the RASSBSC ($92\%$
of the sky). 
The sources are spread over approximately one half of the sky.}
\end{figure}

This fraction can be significantly reduced by removing those 
matches for which the distance between the radio and X-ray positions 
is much larger than the combined NVSS/RASSBSC positional error (when 
this is significantly smaller than the fixed 0.8 arcminutes radius used 
for the cross-correlation).
For a radio/X-ray position distance that is 2.5 times larger than the 
($1 \sigma$) expected value this operation removes 412 entries, a number 
that is close to that expected from spurious associations (see Tab. 1).
We conclude that the remaining (3093) matches should include only a very small 
percentage ($ \lsim 1\%$) of spurious associations. 
Finally, since we are only interested in high galactic latitude, 
point-like, sources we removed all RASSBSC entries with extension larger 
than 30 arcseconds (e.g., Fisher et al. 1998) and with $|b| < 20^{\circ}$. 
The final sample includes 2166 objects.

Our sample of NVSS/RASS sources includes previously known astronomical
objects. These have been identified by cross-correlating the sample with
several astronomical catalogs available within BROWSE (e.g. AGN, HD, SAO, 
VSTARS, CVCAT, EMSS, WGACAT, etc.) and by using NED. 
This operation led to the identification of over 1000 previously known
sources. Optical magnitudes and other parameters for all the identified objects have been added to our database using the values listed in the 
catalogs or provided by NED.
For the remaining sources we have attempted to obtain an
estimate of their optical magnitude by means of the COSMOS (Yentis et al. 
1992) 
and the APM (Irwin, Maddox, \& McMahon 1994) on-line services.
This was achieved through a procedure that automatically retrieves, from the on-line services, a list of potential optical counterparts lying
within 9 arcseconds from the more accurate NVSS position. In most cases only 
one 
candidate optical counterpart is located within the combined 
optical-NVSS uncertainty region, which is typically 5 arcseconds in 
radius. When this occurs we assume
that the only candidate present is the optical counterpart of the
radio/X-ray source. When more than one optical object is inside
in the combined optical/NVSS error region we assume that the optical
counterpart is the brightest source. 

To take into account the fact that the uncertainties in the radio position 
of the NVSS increase at smaller fluxes, we have repeated the COSMOS/APM 
search for the objects without optical counterpart using a 15 arcseconds 
radius. This has resulted in several additional radio/optical associations, 
reducing to only 92 (out of a total of 2166) 
the number of cases where no optical counterpart could be found. 
Some of these could be the result of accidental radio-X-ray 
associations, which are expected to be present at the $\sim 1$ percent 
level; other can be  associated to problematic optical plate regions (e.g., 
plate impurities, regions close to very bright stars, etc.; see below for 
the case of the HBL sample). 
In other cases the optical counterpart could indeed be very faint and below the plates sensitivity limit. 

V magnitudes where estimated from $O$ and $E$ magnitudes obtained 
from the APM for the northern hemisphere and from the COSMOS $B_J$ 
magnitudes as follows.
When $O-E$ was available (APM) we converted $O$ magnitudes first to $B$
magnitudes using $B = O - 0.119\times (O - E)$ (Ciliegi et al. 1996). We 
then derived
$B-R$ from $O-E$ following Gregg et al. (1996), and $B-V$ from $B-R$ using the
standard formulae given by Zombeck (1990; when $O-E$ was not available 
we have simply assumed $O-V=0.5$). Finally, $V = B - (B-V)$.  In the
case of the COSMOS $B_J$ magnitudes, where no color information is given, we
simply assumed $B-V = 0.5$. Galactic extinction ($A_V$) was estimated 
following Wilkes et al. (1994) from the value of the hydrogen column density 
along the line of sight in units of atoms~cm$^{-2}$, $N_H$, derived from the 
Dickey and Lockman (1990) 21cm radio data. 
(Namely, $A_V=3\times[{\rm max}(0,(N_H\times1.99\times 10^{-22}-0.055)]$.)  
We have thus been able to obtain reddening-corrected magnitude estimates for 
2074 sources at $|b|>20^{\circ}$. 

External accuracy for the APM magnitudes is estimated to be $\sim 0.3$
magnitudes for the fainter images, reaching $\sim 1$ magnitude for bright
images. We assume a value of 0.5 but note that for our purposes this is a
conservative estimate. The sources we are interested in, in fact, have
$\alpha_{\rm ro} > 0.2$ (see below) and are therefore relatively faint in the
optical. 

We note that the completeness limits of the plates which were scanned to 
produce the APM and COSMOS catalogs are $O \sim 21.5$ and $B_J \sim 22.5$ for 
the northern and southern sky respectively (Irwin et al. 1994; see also the 
on-line description of the APM catalogues at 
{\it http://www.ast.cam.ac.uk/$\sim$apmcat/}). This translates to $V_{\rm lim} \sim
21$ and 22. 

\subsection{The $\alpha_{\rm ox}-\alpha_{\rm ro}$ diagram} 

The RASSBSC catalog includes all RASS sources detected at a count rate
larger than 0.05 cts/s (although it is only complete down to 0.1
cts/s). This count rate translates into different X-ray flux limits
depending on the amount of $N_H$ present along the line of sight and
on the assumed sources' spectral shape. For our database we calculate
X-ray fluxes assuming a power law spectrum with energy slope
$\alpha_{\rm x} = 1.1$ (a value that is appropriate for high \fxfr BL
Lacs, Padovani \& Giommi 1996) and $N_H$ as derived from the Dickey
and Lockman (1990) 21cm radio data.

With the inclusion of X-ray fluxes our multi-frequency database holds
information on radio, optical and X-ray fluxes for a large fraction of the
NVSS/RASSBSC sources. We have used these data to calculate various
derived quantities. 

\section{Filling the multi-dimensional parameter space}
\begin{figure}
\epsfig{figure=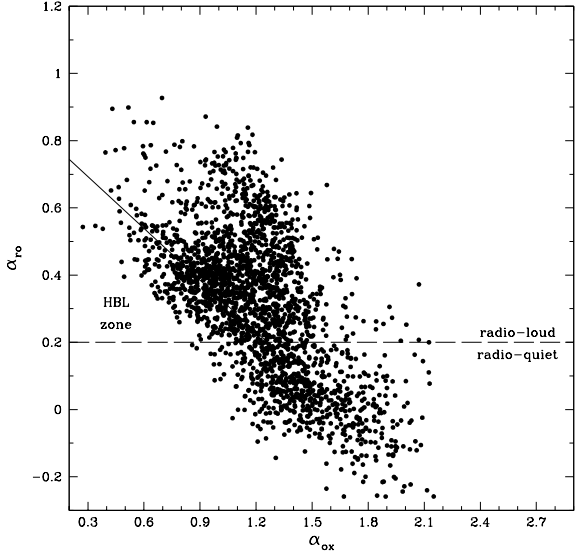,height=8.cm,width=8.0cm}
\caption{The 2074 high-latitude objects in our sample of NVSS/RASSBSC 
sources for which an estimate of the optical magnitude is available are 
plotted in the $\alpha_{\rm ox}-\alpha_{\rm ro}$ plane. The dashed line at
$\alpha_{\rm ro}=0.2$ is our assumed division between radio-loud and 
radio-quiet AGN. Very few BL Lacs fall below this line. The solid line
represents the locus of $\alpha_{\rm rx} = 0.56$, which is equivalent
to an X-ray-to-radio flux ratio \fxfr $= 3 \times 10^{-10}$ \ergj. The 
triangular region delimited by the two lines (HBL zone) includes mostly
BL Lacs (see text for details). For ease of comparison the 
scale is the same as Fig. 3.}
\end{figure}

\begin{figure}
\epsfig{figure=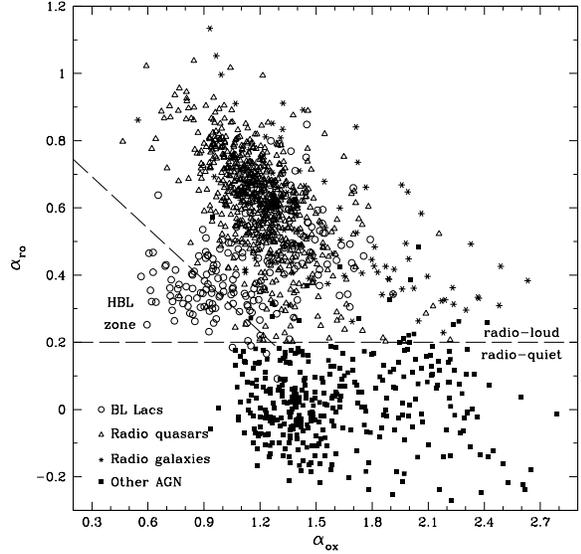,height=8.cm,width=8.0cm}
\caption{The $\alpha_{\rm ox}-\alpha_{\rm ro}$ diagram of  
1362 AGN for which we could find radio, optical, and X-ray data
from the literature. These come from the multifrequency AGN catalogue based 
on  
V\'eron-Cetty \& V\'eron (1996) and described by Padovani, Giommi \& Fiore 
(1997b).
Open circles represent BL Lacs; open triangles are radio quasars; asterisks
are radio galaxies, and filled squares are other types of emission lines
AGN. The dashed line at $\alpha_{\rm ro}=0.2$ is our assumed division between
radio-loud and radio-quiet AGN. Very few BL Lacs fall below this line. The
solid line represents the locus of $\alpha_{\rm rx} = 0.56$, which is
equivalent to an X-ray-to-radio flux ratio \fxfr $= 3 \times 10^{-10}$
\ergj. The triangular region delimited by the two lines (HBL zone) includes 
mostly BL Lacs (see text for details).}
\end{figure}

Figure 2 shows the 2074 sources for which we have radio, optical and X-ray
fluxes plotted in the $\alpha_{\rm ox}-\alpha_{\rm ro}$ plane. These are the
usual effective spectral indices defined between the rest-frame frequencies 
of 5 GHz, 5000 \AA, and 1 keV. X-ray and optical fluxes have been corrected
for Galactic absorption. The k-correction for sources without redshift
information has been derived by assuming a redshift typical for the \fxfr
value of the object (namely, $z = 1$, 0.5 and 0.25 for \fxfr $< 10^{-12}$,
$10^{-12} < $ \fxfr $< 3 \times 10^{-11}$, and \fxfr $> 3 \times 10^{-11}$
\ergj respectively, as derived from the known objects in our database). 
The NVSS 
radio fluxes (at 1.4 GHz) were converted to  
fluxes at 5 GHz by assuming $\alpha_{\rm r} = 0.2$, a value which is 
appropriate for HBLs. Similarly, $\alpha_{\rm o} = 0.65$ was assumed (Falomo,
Scarpa \& Bersanelli 1994). 

Uncertainties on the $\alpha_{\rm ox}$, $\alpha_{\rm ro}$, and \fxfr values
will arise because of the non-simultaneity of the flux measurements and
because of intrinsic uncertainties. Large flux variations, especially in the 
optical and X-ray band, can lead to $\Delta \alpha_{\rm ox}$ up to 0.2-0.3  
and to $\Delta \alpha_{\rm ro} \lsim 0.15$
Uncertainties due to flux measurements are expected to be smaller since 
NVSS radio fluxes
are typically good to better than $\sim 10\%$ and RASSBSC count rates have
on average 20\% uncertainties. Adding a 0.5 magnitude error in the optical band
(see previous subsection) gives these typical uncertainties: 
$\Delta \alpha_{\rm ox} \sim 0.08$, $\Delta \alpha_{\rm ro} \lsim 0.04$, and
$\Delta$\fxfr $\lsim 22\%$. 

No large previously unpopulated regions of the $\alpha_{\rm ox}- 
\alpha_{\rm ro}$
plane are filled by our NVSS/RASSBSC sources in the region which is defined by
our survey limits (i.e., $f_{\rm x} \gsim 6 \times 10^{-13}$ \ergs, 
$f_{\rm r} > 2.5$
mJy, $V \lsim 21 - 22$). This indicates that if new source types showing
peculiar and previously unknown broad-band spectrum exist, their number must 
be small. Given the dynamical range of the two surveys we used, it is however
apparent that our selection method favors objects towards the lower left
corner of the $\alpha_{\rm ox}-\alpha_{\rm ro}$ plane (compare Fig. 2 and 3). 

The maximum density of sources in Fig. 2 is located roughly along a band of
$\alpha_{\rm ro} \sim 0.2-0.4$ (which normally includes BL Lacs and flat
spectrum radio quasars) and in a second region delimited 
by $1.1 \lsim \alpha_{\rm ox}
\lsim 1.8$ and $-0.1 \lsim \alpha_{\rm ro} \lsim 0.2$ which is populated by 
radio-quiet AGN. The diagonal branch going from $\alpha_{\rm ox} \sim 1.4,
\alpha_{\rm ro} \sim 0.3$ to $\alpha_{\rm ox} \sim 1.0, \alpha_{\rm ro} 
\sim 0.8$, normally populated by bright flat spectrum radio quasars and radio 
selected BL Lacs (LBLs), is also visible but not very pronounced due to the
small number of objects with very bright radio flux and relatively low X-ray
flux (i.e., most of the objects in this area would be too faint to be in the
RASSBSC for a radio flux that is $\lsim$ 1 Jy).

\section{Building a large sample of HBL BL Lacs}

Our goal here is to statistically select a large sample of HBL BL
Lacs by exploiting the fact that this type of sources exhibit a peculiar and
nearly unique radio-through-X-ray spectral energy distribution. It is well
known (e.g., Stocke et al. 1988, 1991, Padovani, Giommi \& Fiore 1997a) that 
there is a region in the $\alpha_{\rm ox}-\alpha_{\rm ro}$ plane that is almost
exclusively populated by HBL BL Lac objects. 
Figure 3 shows the \aoxaro~plot
for 1362 known AGN with radio, optical, and X-ray data, 
taken from various catalogs. These come from the multifrequency AGN catalogue 
based on  V\'eron-Cetty \& V\'eron (1996) and described by Padovani, Giommi \& 
Fiore (1997b). We can define an
area in the figure which contains a very large fraction of BL Lacs, in
particular BL Lacs with a synchrotron peak at UV/X-ray energies or HBL BL
Lacs (given their relatively low values of $\alpha_{\rm ro}$ and $\alpha_{\rm
ox}$: Padovani \& Giommi 1995b). BL Lacs make up $\sim 91\%$ of all AGN in the
triangular area delimited by $\alpha_{\rm ro} > 0.2$ (dashed line) and \fxfr
$\gsim 3 \times 10^{-10}$ \ergj\footnote{The units of this ratio are certainly 
``non-standard'' but those of the numerator and denominator are, so that 
this ratio is easily translated in terms of ``observed'' quantities. 
$\alpha_{\rm rx}$, on the other hand, requires redshift information, not
available for most of our objects, and 1 keV fluxes, which are more dependent
on the precise value of the X-ray spectral index than broad-band X-ray 
fluxes.} (or $\alpha_{\rm rx} \lsim 0.56$: solid line), with
$f_{\rm x}$ in the $0.3 - 3.5$ keV band and $f_{\rm r}$ at 5 GHz. (This
\fxfr cut is a compromise between sample size and percentage of BL Lacs). 
We will call this area {\it the HBL zone}. 
It must be noted that these statements are strictly valid only in the flux 
range covered by the surveys used to build Figure 3. At fainter flux 
levels new populations of sources could emerge changing the simple picture
described above.
The X-ray, radio and optical fluxes of our NVSS/RASSBSC sources is well 
within the region of the parameter space covered by several other surveys, 
(e.g. the {\it Einstein} EMSS, the IPC Slew Survey, the EXOSAT HGLS and 
the HEAO1 surveys) so we feel that we can proceed safely. 

For spectral indices typical of
HBLs, it turns out that $f_{\rm x}(0.3 - 3.5~{\rm keV})/f_{\rm r}(5~{\rm GHz})
\sim 1.1~f_{\rm x}(0.1 - 2.4~{\rm keV})/f_{\rm r}(1.4~{\rm GHz})$. From now 
on, all \fxfr ratios will refer to the {\it ROSAT} and NVSS bands.

For each unidentified object in the HBL zone in our multifrequency database
we have retrieved 
an optical image from the on-line Digitized Sky Survey (DSS) service of the 
Space Telescope Science Institute 
(http://archive.stsci.edu/dss) and we have visually inspected it after 
plotting 
both the X-ray and radio error circles to search for 
obvious problems. 
In a few cases we found that the X-ray/radio association was clearly spurious 
due to the presence of a bright star ($V$ \lsim 10) non-coincident 
with the 
NVSS source. The star is most probably the counterpart of the X-ray emission. 
In other cases the RASS error box included a relatively bright galaxy 
($V\sim 15-16$), presumably the source of the X-ray emission, and the 
position of the radio source was clearly inconsistent with that of the 
extended object. All these sources were removed from the sample. 

After these checks our multifrequency database includes 234 sources in the HBL
zone. {}From a cross-correlation with catalogs of known objects we see that
74 objects were already classified, 60 of these being previously known BL
Lacs. Three other sources are galaxies in clusters (these could be BL Lacs in
clusters, although the cluster gas emission could also provide the X-rays
artificially moving the galaxy into the HBL zone), one is a galaxy listed in the
CfA Redshift Catalog (Huchra et al. 1992), for which we have a redshift but no
other information (its X-ray luminosity, however, is $L_{\rm x} \sim 3 \times
10^{44}$ erg s$^{-1}$, implying an active nucleus). In 10 cases our candidate
HBLs are instead coincident with known emission line AGN. We
have inspected the published optical spectra of these sources to check whether
they could be misclassified BL Lacs. In all cases where this was possible
(7/10) we have found 
strong and broad emission lines, typical of Seyfert galaxies or QSO.  
The presence of these sources (all of them close to the borders of the HBL 
zone, both in terms of $\alpha_{\rm ro} [\sim 0.25]$ and $\alpha_{\rm rx} 
[\sim 0.52]$) is probably due to one of the following reasons:
1. uncertainties in the estimation of the optical magnitude, which produces a
scatter $\Delta\alpha_{\rm ro}$ $\sim 0.08\Delta V$ ($\Delta V$ expressed in
magnitudes); 2. large X-ray flares from quasars close to the borderline area;
3. clusters/groups of galaxies with size $\lsim 30$ arcseconds; 4. radio-loud 
AGN with multifrequency spectra similar to those of HBLs (Padovani et al. 
1997b; Perlman et al. 1998). These 10 sources will not be considered for the 
estimation of the \vovm and logN-logS of HBLs.  

As an additional criterion to reduce the contamination from non BL Lacs
we excluded sources for which the RASS hardness ratio HR1 [defined as 
$(cts_{0.5-2.0keV}-cts_{0.1-0.4keV})/(cts_{0.5-2.0keV}+cts_{0.1-0.4keV})$] 
is softer
than $-0.5$ (see Cao et al. 1998) thus removing sources with strong
soft excess. This method can be efficiently applied to our extreme 
HBL candidates
since their X-ray spectra are expected to be relatively flat 
($\alpha_x \lsim 1.5$; see Fig. 2 of Padovani \& Giommi 1996), but not to less 
extreme (intermediate) BL Lacs which can be as steep as $\alpha_x \sim 3$. 
Within our sample only 8 sources have HR1 $< -0.5$. Six of these are 
unclassified but the two known sources are both emission line AGN. 

The percentage of confirmed BL Lacs among the previously 
identified sources in our sample ranges between 83 and 88\% depending 
on whether the 
galaxies in clusters are actually BL Lacs or spurious associations. 
This fraction grows to $ 95-100\% $ by imposing the more restrictive limits 
\fxfr $ > 5 \times 10^{-10}$ and $\alpha_{\rm ro} > 0.25$ but at the cost of
nearly halving the sample. This effect could simply be due 
to the fact that the percentual errors
in the flux values become larger near the survey limits, causing some
neighbouring objects to cross the border. In fact, the 
fraction of BL Lacs becomes 95\% for $f_{\rm x} > 5\times 10^{-12}$ \ergs and
100\% for $f_{\rm x} > 1\times 10^{-11}$ \ergs. It is difficult however, 
to distinguish
between the effect of larger errors at faint fluxes and a real change
in the underlying population as a function of flux. If there is such 
a population change within the flux range considered in this work it 
must be limited to a small percentage. 

The statistical identification method described in this paper could 
also be applied to samples with somewhat less extreme values of \fxfr than 
those considered here. 
In this case the efficiency decreases but not very rapidly. For example, 
a cut at \fxfr $\ge 3 \times 10^{-11}$ 
would result in a sample in which $\sim 61\%$ of identified objects 
are BL Lacs. 

\subsection{A well-defined radio flux limited sample of HBL BL Lacs} 

The sample of 218 sources described in the previous paragraph includes 
a very high percentage of HBL BL Lacs and therefore can be 
used to select a very large sample of HBL BL Lacs. This sample is 
much larger than the presently available X-ray selected samples 
which include typically only $30-40$ objects. 
The definition criteria used however are not strict enough to allow 
detailed statistical analysis. 
In this paragraph we define a flux-limited subsample that can be used for 
statistical purposes. 
This restricted sample is defined by the following conditions:

\begin{enumerate}
\item \fxfr $\ge 3\times 10^{-10}$ \ergj;
\item $f_{\rm r} \ge 3.5$ mJy;
\item RASSBSC count rate $\ge 0.1$ cts/s;
\item $\alpha_{\rm ro} > 0.2$;
\item $V \le 21$;
\item $|b|>20^{\circ}$;
\end{enumerate} 

Conditions i) and ii) make this a radio flux limited sample of extreme
HBL BL Lacs. The resulting limiting X-ray flux is in fact $f_{\rm x}
\ge 1.05 \times 10^{-12}$ \ergs and there are no sources in our sample with
\fxfr $\ge 3\times 10^{-10}$ \ergj and count rate $\ge 0.1$ cts/s 
[condition iii)] below this limit. In other words, our survey has detected all
objects with \fxfr $\ge 3\times 10^{-10}$ \ergj and $f_{\rm r} \ge 3.5$ mJy in
the area we cover (see next section). (Note that we {\it have not} detected
all objects with \fxfr $\ge 3\times 10^{-10}$ \ergj and $f_{\rm x} \ge 1.05
\times 10^{-12}$ \ergs, as we miss all sources with $f_{\rm r} < 3.5$ mJy.
Our survey then is not complete for extreme HBLs in the X-ray band.)
Condition iii) ensures that the coverage of the RASSBSC catalog ($92\%$ of the
sky) also applies to our sample. Condition iv) (which excludes ``radio-quiet''
sources) makes sure that our candidates are fully within the HBL
zone. Relaxing this condition would include a few more BL Lacs but at
the expenses of a large contamination from other AGN classes just below the
dividing line (see Figs. 2 and 3). The optical magnitudes of sources
satisfying condition i -- iv can reach values near 21 (our assumed
completeness limit in the optical band) only in a very restricted corner of
the parameter space, namely for $\alpha_{\rm ro} \sim 0.5$ and $f_{\rm r}$
just above 3.5 mJy. For this reason we have added condition v), which makes
the sample both radio and optically flux-limited, although the latter is a
very mild limit. [We stress that only 2 empty fields satisfy conditions 
i), ii), iii), and vi). These fields remain ``empty'' even in the second
generation scans of the DSS, which are deeper than the first one. This shows
that the optical counterparts are safely below are assumed value of 
$V_{\rm lim} = 21$.]
Finally condition vi) simply restricts our sample to high
galactic latitude sky regions.
Conditions i) through vi) applied to our multi-frequency database 
define a set of 155 objects.  
The main steps that have led to this final sample are summarized in 
Table 2.
A dedicated paper presenting finding charts, X-ray, optical, and radio
fluxes for these sources is in preparation and will be published in the
near future. 
Although our sample is far from being
completely identified, it currently includes 58 BL Lacs and 
its HBL content is expected to be $\sim 85\%$. 
Taking this limitation in mind in the following sections we use this 
well-defined sample to investigate the statistical properties of HBL BL Lacs.

\begin{table*}
\begin{center}
\begin{tabular}{|l|l|c|}
\multicolumn{3}{c}{Table~2: Main steps followed for the selection of the complete HBL sample}\\ \\
\hline \\
Number of &Operation & No. of remaining \\
objects & & objects\\
&&\\ \\[-3mm]
\hline \\[-3mm]
1,807,316(NVSS) & Cross-correlation of NVSS and RASSBSC catalogues  & --- \\
~~~~18,811(RASSBSC) & (correlation radius=0.8 arcminutes) & 3,505 \\
~~~~~3,505&Removal of matches with $\Delta$ position (radio/X-ray) $ > 2.5 \sigma $&3,093 \\
~~~~~3,093&Restrict to point-like (X-ray size $\le 30^{\prime\prime}$) and 
high Galactic latitude sources ($|b| > 20^{\circ}$) & 2,166\\
~~~~~2,166&Take sources with optical counterpart in APM/COSMOS            & 2,074\\
~~~~~2,074&Take only sources in the HBL zone: \aro $>$ 0.2, \fxfr $> 3\times 10^{-10}$ \ergj & ~~234\\
~~~~~~~234 &Removal of known broad line objects & ~~224 \\
~~~~~~~224 &Removal of sources with ROSAT hardness ratio HR1 $<-0.5$  & ~~218\\
~~~~~~~218&Further conditions for a complete sample: ROSAT cts/s $>$ 0.1, $f_{\rm r} > 3.5$ mJy, $V \le 21$  & ~~155\\
\hline
\end{tabular} 
\end{center}
\end{table*}

\subsection{The sky coverage of the sedentary survey} 

The sky coverage of our survey is given by the coverage of the NVSS (that is
$\delta > -40^{\circ}$), the galactic latitude cut ($|b|> 20^{\circ}$), and by
the sky coverage of the RASSBSC. The first two limits imply a survey area of
22,990 deg$^2$. The current, practically final, version of the NVSS includes
7,432 more sources than the one we used, which means we are roughly missing
$0.4\%$ of that area, for a total of 22,896 deg$^2$. The RASSBSC covers $92\%$
of the sky at a limit of 0.1 cts/s (Voges et al. 1996) which is also the limit
that we use (condition iii) in the definition of the HBL sample. 
The sky coverage of the RASSBSC, however is not simply a flat 92 percent 
of the sky since the varying amounts of $N_H$ in different directions make 
the sensitivity a function of right ascension and declination. 
Assuming a power law spectrum with energy index of 1.1 and taking into account 
the absorption due to Galactic $N_H$ (from the 21cm measurements of Dickey 
and Lockman 1990)  for sources with $\delta > -40^{\circ}$ and count rate 
$\ge 0.1$ cts/s, we obtained the sky coverage shown in Fig. 4.
The combined NVSS/RASSBSC sky coverage then amounts to a total area of 
$21,064$ deg$^2$ with a dependence on X-ray flux shown in Fig. 4. 

\begin{figure}
\epsfig{figure=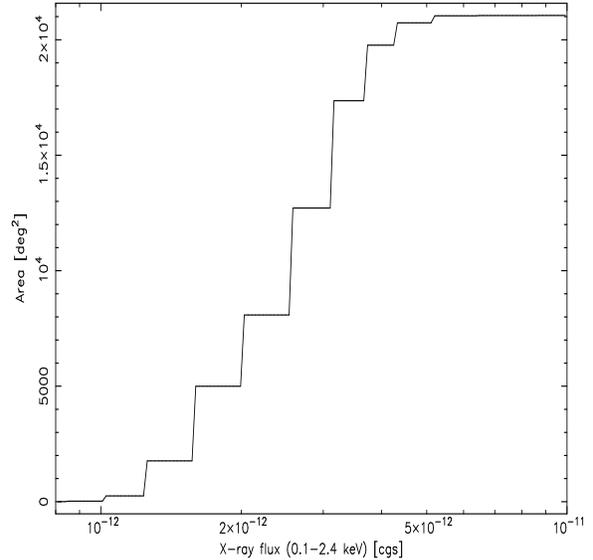,height=8.5cm,width=8.5cm,angle=-90}
\caption{The sky coverage of the RASSBSC catalog for sources with 
$\delta > -40^{\circ}$ and count rate in excess of 0.1 cts/s. The 0.1-2.4 keV
flux has been calculated assuming a power law spectrum with energy index
$\alpha =1.1$ and absorption due to Galactic $N_H$ as estimated from 21cm data.}  
\end{figure}

\section{The radio LogN-LogS and \vovm distribution of HBL BL Lacs} 

The cosmological properties of X-ray selected BL Lacs (mostly HBLs, Giommi, 
Ansari \& Micol 1995) have 
been reported as being different from those of all other types of AGN 
(Maccacaro et al. 1988, Morris et al. 1991, Wolter et al. 1994, Bade et al. 
1998). Namely, it has been suggested that HBLs might be less luminous
and/or less numerous at high redshift, a behavior which is unique amongst
extragalactic sources. To test if this result is also present in a radio flux
limited sample we have calculated the LogN-LogS and the \vovm
distribution (Schmidt 1968), or more properly \vova
(Avni \& Bachall 1980) for the sample defined in the previous section.
The integral LogN-LogS of this sample of HBLs is shown in Fig. 5 together 
with a Euclidean power law with slope $-1.5$. 
A flattening at around $\sim 20$ mJy is evident from the plot. 
In the same figure we also plot the radio counts for the 8 EMSS HBLs in the 
Morris et al. (1991) sample (as updated by Rector, Stocke \& Perlman 1999)
with \fxfr $ > 3\times 10^{-10}$ and $f_{\rm r} > 1.7$ mJy 
(33\% of the revised sample of 24 objects). 
In analogy to our sample, we can in fact 
derive a radio flux-limited subset of the EMSS BL Lacs. For an X-ray limit
of $5 \times 10^{-13}$ \ergs, the condition \fxfr $ > 3\times 
10^{-10}$ translates into a radio flux limit of 1.7 mJy. (In other words, 
despite being an X-ray flux limited sample, the EMSS could detect in the radio 
band all BL Lacs with \fxfr $ > 3\times 10^{-10}$ above $f_{\rm r} \sim 1.7$
mJy.) Only one of 
these extreme EMSS BL Lacs has a radio flux below this value and was therefore
excluded. The remaining 8 objects make up a
complete, albeit small, radio sample which can be compared to ours. 
The radio counts of these objects are in good agreement with ours, within
their (rather large) uncertainties. 

We also show the expected radio counts for all types of BL Lacs (top
line) estimated from the radio luminosity function of Padovani \&
Giommi (1995b) that it is based on a beaming model and the radio luminosity 
function of Fanaroff-Riley type I radio galaxies (Urry et al. 1991),
assumes no evolution, and fits the luminosity function of the 1 Jy BL
Lacs.  Note that the number of HBLs with \fxfr $> 3\times 10^{-10}$
\ergj~is a roughly constant fraction ($\sim 1/50$) of the surface
density of the whole population of BL Lacs.

We have next estimated the \vova distribution of the HBLs in our sample. 
The exact value of \vova of a source 
depends somewhat on its redshift thus introducing uncertainties in the 
\vova distribution when $z$ is not known and an estimate has to be 
used in its place. (For example, for $f/f_{\rm lim} = 2$, \vova ranges 
between $0.35$ and $0.6$ for $z$ which goes between 0 and 1.5. 
The dependence is weaker for lower values of $f/f_{\rm lim}$.)
For all unidentified sources and for those objects identified with catalogued
BL Lacs for which a redshift is not available we have assumed $z = 0.25$, a
value close to the average redshift in the subsample of identified BL Lacs.
We obtained \vovaave $ = 0.42 \pm 0.02$ implying that our radio flux limited
sample of HBLs shows a significant deviation ($\sim 4\sigma$, $P > 99.9\%$) 
from a population of constant cosmological density.

To investigate the dependence of this result on redshift we have then
calculated the \vova distribution assuming $z=0.5$ for the sources without
redshift, a value which is more than a factor of 2 larger than the average
redshift of the identified BL Lacs in the sample. The result is \vovaave $=
0.45 \pm 0.02$, still more than $2 \sigma$ lower than 0.5.  As a comparative
check we have also repeated the test on a sample selected imposing the same
definition criteria used for the HBL sample except for point (iii) which has
been changed to $\alpha_{\rm ro} < 0.15$. This new sample of radio quiet
objects ($\alpha_{\rm ro} < 0.15$) includes 229 sources. Of these, 157 are
identified, as follows: 14 stars (with relatively low $\alpha_{\rm ro}$
values), 13 galaxies, 2 BL Lacs, 1 Liner, 4 Seyfert 2s, and 123 broad-line
objects (Seyfert 1 and quasars). The average redshift of the identified
(extra-galactic) objects is 0.09. Most of the remaining unidentified sources
are then expected to be nearby type 1 AGN. The radio flux in these sources is
not necessarily directly related to the X-ray flux which is produced by a
mechanism other then Synchrotron emission. The result is \vovaave $= 0.53 \pm
0.02$ (for unidentified objects we have assumed $z=0.09$, equal to the average
value seen in the identified sources) consistent with a population of objects
characterized by no cosmological evolution. The LogN-LogS function for this
control sample is shown in Fig. 6, where no deviation from a Euclidean power
law with slope $-1.5$ (dotted line) is apparent.

The expected contamination in the HBL sample due to the spurious presence of 
$\lsim 15\%$ of emission line AGN should therefore 
increase the value of \vovaave  since the measured \vovaave of 
these objects is larger than that of HBLs. The final value of \vovaave for 
our HBLs might therefore be even less than 0.42 once the sample is 
completely identified.  
We conclude that the flat LogN-LogS of HBLs and their
low value of \vovaave is intrinsic to HBLs and not due to 
some unknown bias introduced by our selection method.

We have next calculated \vovaave for different values of the radio 
flux limit, keeping unaltered the other conditions detailed in Sect. 4.1. 
The results are reported in Table 3 for HBLs and Table 4
for the control sample of radio quiet objects.

\begin{table}
\begin{center}
\begin{tabular}{|c|c|c|}
\multicolumn{3}{l}{Table~3: \vovaave of extreme HBLs for different }\\
\multicolumn{3}{l}{radio flux limits }\\ \\
\hline \\
Radio flux limit & \vovaave & No of objects \\
mJy & &in sample\\ \\[-3mm]
\hline \\[-3mm]
 3.5 & $0.42\pm 0.02$ & 155 \\
 5.0 & $0.43\pm 0.02$ & 135 \\
 ~10 & $0.45\pm 0.03$ & ~84  \\
 ~20 & $0.49\pm 0.04$ & ~44  \\
\hline
\end{tabular} 
\end{center}
\end{table}

\begin{table}
\begin{center}
\begin{tabular}{|c|c|c|}
\multicolumn{3}{l}{Table~4:  \vovaave of radio quiet AGN for different }\\
\multicolumn{3}{l}{radio flux limits }\\ \\
\hline \\
Radio flux limit & \vovaave & No of objects \\
mJy & &in sample \\ \\[-3mm]
\hline \\[-3mm]
 3.5 & $0.53\pm 0.02$ & 229 \\
 5.0 & $0.51\pm 0.02$ & 145 \\
 ~10 & $0.49\pm 0.04$ & ~59  \\
 ~20 & $0.50\pm 0.06$ & ~20  \\
\hline
\end{tabular} 
\end{center}
\end{table}

\section{Selection effects}

\begin{figure}
\epsfig{figure=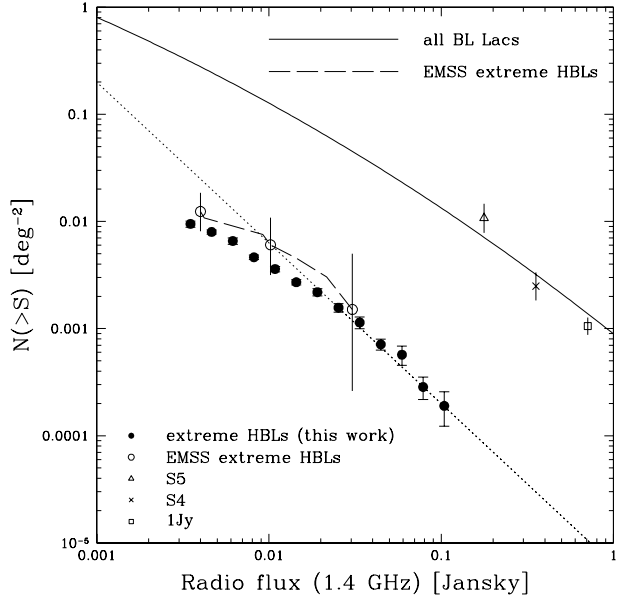,height=8.5cm,width=8.5cm}
\caption{The radio LogN-LogS of the HBL sample (\fxfr $ > 3\times 10^{-10}$; 
filled
circles). A flattening is apparent at around 20 mJy. The dotted line 
represents the Euclidean slope of $-1.5$. 
The radio counts of the 8 EMSS HBLs in the Morris et al. (1991) sample 
with \fxfr $ > 3\times 10^{-10}$ and $f_{\rm r} >1.7$ mJy are also shown 
(dashed line and open 
circles; see text for details); for clarity, error bars are given only for 
3 representative points. 
The solid line 
represents the expected radio counts for all types of BL Lacs estimated 
from the radio luminosity function of Padovani \& Giommi (1995b). The 
BL Lac surface density from the 1 Jy (open square), S4 (cross) and S5 
(open triangle) 
surveys (the latter two still not completely identified), in excellent 
agreement 
with the prediction, are also shown. All data apart from our own HBLs
(filled circles) have been converted from 5 GHz, assuming $\alpha_{\rm r} 
=-0.27$ for the 1 Jy, S4, S5 points and beaming prediction and  
$\alpha_{\rm r} = 0.2$ for the EMSS HBLs.} 
\end{figure}

\begin{figure}
\epsfig{figure=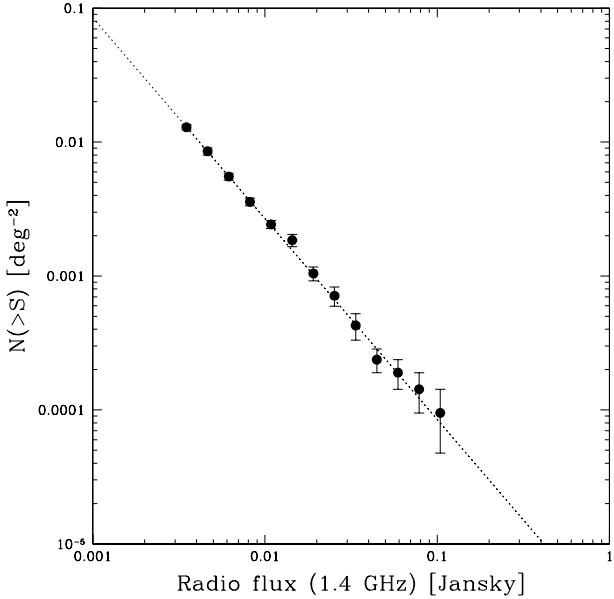,height=8.5cm,width=8.5cm}
\caption{The LogN-LogS of the control sample of 229 radio quiet AGN with 
\fxfr $ > 3\times 10^{-10}$. The counts are consistent with the Euclidean 
slope of $-1.5$ (dotted line)}
\end{figure}

The sample selected on the basis of \fxfr $ > 3 \times 10^{-10}$
\ergj~with the further restriction of $\alpha_{\rm ro} > 0.2$
guarantees a very high fraction ($\sim 85\%$) of HBL BL Lacs without
the need of spectroscopic classification that in some cases may
introduce selection effects. One of these arises from the (4,000 \AA )
Ca II break in the optical spectrum of a candidate which is used to
classify the object as a BL Lac or as a radio galaxy otherwise (e.g.,
Stocke et al. 1991, Perlman et al. 1998, Rector, Stocke and Perlman,
1999). Our method is independent of this and in principle can measure
its reliability in distinguishing radio galaxies from BL Lacs once our
sample is completely identified from follow-up optical
spectroscopy. We note however that the contribution from the host
galaxy in the optical band flattens the radio to optical index
($\alpha_{\rm ro}$) and could push a few BL Lacs with values close to 
our assumed limit ($\alpha_{\rm ro}=0.2$) into the radio-quiet AGN zone (see
Fig. 3). Only four such objects are present amongst the 62 previously
known BL Lacs with \fxfr $ > 3 \times 10^{-10}$ \ergj; their \vovaave
value ($0.34$) is below average and their addition to the sample of
155 has a negligible impact on the radio counts and \vovaave. Such a
loss is numerically compensated by the residual percentage
($\sim 1 \%$) of spurious radio/X-ray association (see Sect. 2), and 
by the small percentage of non BL Lacs expected to be present among the 
93 unidentified objects in the HBL sample. 
As mentioned above, the contamination 
due to the spurious presence of $\sim 10\% $ of emission
line AGN, should increase the value of \vovaave since the measured
\vovaave of these objects is $0.53 \pm 0.02$.

A somewhat related issue is that of the so-called Browne \& March\~a
effect. Browne \& March\~a (1993) suggested that optically faint BL Lacs might
go unrecognized as they could be swamped by the light of the host galaxy,
which is typically a bright elliptical. However, we do not think that this has
much influence on our HBL sample, as our X-ray flux limit is relatively high,
above the value below which this effect starts to become severe (see Fig. 5 of
Browne \& March\~a 1993)\footnote{Note that although our sample is
radio-selected, we have to use the Browne \& March\~a predictions for the
X-ray case, as they use an $\alpha_{\rm ox}$ distribution which is similar to
that of our sources, all HBLs.}. One could argue that such objects, if they
existed, might have $\alpha_{\rm ro} < 0.2$ due to the galaxy component. We
then looked for previously known objects without strong, broad lines, 
with relatively
high radio power ($P > 3 \times 10^{23}$ W Hz$^{-1}$, typical of 
Fanaroff-Riley type I radio galaxies), and $0 <
\alpha_{\rm ro} < 0.2$ (i.e., with a galaxy component up to 2.5 magnitudes
brighter than the BL Lac component). We found only one, a 15th magnitude CD 
galaxy at $z = 0.111$. 

Finally, we have checked how many good matches could have been contaminated by
the accidental presence of an unrelated optical source in the radio error
region for our sample with \fxfr $ > 3 \times 10^{-10}$ \ergj. To this end we
have re-run the automatic magnitude requests procedure from the COSMOS/APM
services after shifting all radio coordinates by a fixed amount. We found that
only in a tiny number of cases ($\sim 2-3 $ in the entire sample) a bright ($V
< 16$ ) spurious optical source could have moved a good candidate past the
radio quiet line ($\alpha_{\rm ro} < 0.2$). Conversely, a truly empty field
(i.e., no counterpart of the radio source down to V = 21) could have been
included in the sample because of the accidental presence of an unrelated
optical source only in less than 10\% of the cases. However, since the number
of empty fields is small this contamination is expected to be negligible.

\section{Discussion and conclusions}

By cross-correlating public catalogs we have assembled a multi-frequency
database of high Galactic latitude sources which includes 2074 
radio--optical--X-ray sources, $59\%$ of them still unidentified. 
The source distribution in the $\alpha_{\rm ox}-\alpha_{\rm ro}$ plane 
(Fig. 2)
follows a well known pattern (Stocke et al. 1991, Padovani et al. 1997a) 
with a horizontal band centered around $\alpha_{\rm ro}=0.3-0.4$ 
for $\alpha_{\rm ox}$ values less than $\lsim 1.5$ (probably tracing 
synchrotron emission, Padovani \& Giommi 1995b) and with a diagonal branch 
(probably tracing inverse Compton emission) populated by LBL BL 
Lacs and by Flat Spectrum Radio Quasars (FSRQ). The region above 
$\alpha_{\rm ro}=0.2$ (radio-loud region) and in between the two main branches 
is less populated similarly to the case of Fig. 3 where only 
radio sources known before our sample was assembled are plotted. No new 
large population of objects filling 
previously unpopulated regions of the $\alpha_{\rm ox}-\alpha_{\rm ro}$ plane 
seems 
to emerge. This indicates that the dominant mechanisms producing 
electromagnetic radiation in the full sample of radio-loud objects are 
not different from those seen in known sources, that is Synchrotron and 
Synchrotron/Compton emission.

We have been able to statistically identify a large sample of extreme
(\fxfr $> 3 \times 10^{-10}$ \ergj) HBL BL Lacs with
very small contamination ($\lsim 15\%$) from other sources. We have named this 
survey {\it
sedentary} since it can be used to draw conclusions about some of the
cosmological properties of High Energy Peaked BL Lacs even without complete
optical identification. 
This type of BL Lacs are typically found in X-ray surveys (e.g.,
EMSS, Stocke et al. 1991, EXOSAT HGLS, Giommi et al 1991, \einstein IPC 
Slew survey, Perlman et al. 1996). For the first time we have been 
able to assemble a sizable and well defined sample of HBL BL Lacs 
that is {\it radio} flux 
limited with a sensitivity limit that is nearly three orders of magnitude
lower than the only existing complete sample of BL Lacs 
at radio frequencies (the 1Jy sample, Stickel et al. 1991).

The method presented here could also be extended to select large samples 
including a high percentage ($\sim 30-50\%$) of BL Lacs with less 
extreme HBL properties, although in this case follow up optical 
observations cannot be avoided.

The main result of this study is that our sample of HBLs shows a 
\vovaave $= 0.42 \pm 0.02$. This value is somewhat dependent on the redshift
assumed for the sources without redshift determination but even if this is
taken to be twice as large as $\langle z \rangle \sim 0.25$ \vovaave is
still below 0.5 at the 96\% level. 
The space distribution of these 
objects appears then to be inconsistent with that of uniformly distributed 
population of sources. 
This result  extends to the radio band earlier X-ray findings based on much smaller samples of BL Lacs discovered in 
\einstein and \rosat X-ray images (Maccacaro et al. 1988, Morris et al. 1991, 
Wolter  et al. 1994, Bade et al. 1998). 

It is also interesting that the low ($< 0.5$) \vovaave values set in 
at $f_{\rm r} \lsim 20$ mJy (see Tab. 3), that is around the radio flux 
where the counts start to flatten. Both effects can be explained by a dearth 
of low-flux sources. No such dependence of \vovaave on radio flux is
present for the radio-quiet control sample. 

These results have been interpreted as evidence of a ``negative'' amount 
of cosmological evolution, that is  BL Lacs would be more numerous or 
more luminous now than in the past. 
Padovani \& Giommi (1995b) noticed that this non-uniform distribution 
could be explained by the fact that at higher redshifts, and therefore 
higher rest-frame frequencies, one would be more likely to be above the 
peak of the Synchrotron emission, thereby ``losing'' objects. In other 
words, X-ray selection would preferentially pick-up nearby,
high energy peaked HBL BL Lacs (which are much stronger X-ray emitters than
LBLs), whereas higher luminosity/redshift objects would be more difficult to
detect, causing an apparent lack of X-ray selected BL Lacs at high
redshift. Padovani \& Giommi (1995b), however concluded that at that time 
this hypothesis was not supported by sufficient unbiased experimental evidence.

Another possible reason for this peculiar behavior could be related to 
different evolution rates of the thermal (disk + emission lines) and the 
non-thermal (synchrotron/Compton) components in blazars. 
If the thermal component is characterized by a more pronounced cosmological 
evolution then sources at high redshift would 
be called FSRQ more frequently than at low redshift (or, equivalently, they
would be classified as BL Lacs less frequently than at low redshift).  
One such scenario has been proposed by Cavaliere \& Malquori (1999) where 
the slowly evolving synchrotron/Compton component is produced via the 
Blandford-Znajek (1977) process, characterized by long timescales 
while the QSO/thermal component is instead short lived and is triggered by 
merging events that were more frequent in the past.
This scenario however would require a tight connection between 
BL Lacs and FSRQ, which at the moment does not seem to be supported 
by observations (see, e.g., Urry \& Padovani 1995). 

Fossati et al. (1998), studying flux limited samples of BL Lacs selected at
different frequencies reached the conclusion that synchrotron emission in
powerful BL Lacs peaks at IR frequencies whereas in less luminous objects the
peak is located near the X-ray band. Ghisellini et al. (1998) have interpreted
this as the consequence of an intrinsic correlation between the Lorentz factor
of the electrons emitting near the synchrotron peak and the energy density
present in the emitting region. In this model most low luminosity BL Lacs
would be HBLs and most high luminosity BL Lacs would be LBLs. This is a strong
conclusion that can be tested with our results. The radio LogN-LogS plotted in
Fig. 5 shows that the fraction of extreme HBLs compared to the expectation for
all types of BL Lacs is always small ($\sim 2\%$) and roughly constant down to
the 3.5 mJy limit of our survey. This seems to be inconsistent with the expectations of
the Fossati et al. (1998) and Ghisellini et al. (1998) model that instead
predicts that the percentage of HBL BL Lacs must significantly increase 
with decreasing radio luminosity until it becomes the dominant component 
at low radio fluxes/luminosities. For example, the total HBL fraction 
is predicted to go from $5\%$ at 1 Jy, to $16\%$ at 100 mJy, to $38\%$ 
at 10 mJy, with an overall increase of a factor of 8 (Fossati et al. 1997). 
This implies a steep radio logN-logS of HBLs that is ruled out by the 
actually measured logN-logS shown in Fig. 5. Any steepening must start 
below our survey flux limit.
Two points should however be noted: 1. our survey includes extreme HBLs, 
i.e., objects with X-ray-to-radio flux ratios \fxfr $> 3\times 10^{-10}$ \ergj
(the dividing line between HBLs and LBLs is generally taken to be around
\fxfr $\sim 1\times 10^{-11}$ \ergj; see, e.g., Padovani \& Giommi 1995b). 
These however make up a substantial fraction of all HBLs (8/24 or 
$33\%$ in the revised EMSS sample of Morris et al. 1991); 2. the radio counts 
for the whole BL Lac population are based on the 1 Jy sample and on a 
beaming model (see Padovani \& Giommi 1995b for details). They 
therefore represent our best guess for the number density of BL
Lacs at low radio fluxes but they still are an extrapolation. We note,
however, that this extrapolation is in excellent agreement with the BL Lac
number density derived from the S4 ($f_{\rm 5GHz} \geq 0.5$ Jy; Stickel \&
K\"uhr 1994) and S5 ($f_{\rm 5GHz} \geq 0.25$ Jy; Stickel \& K\"uhr 1996)
catalogs. The comparison will be on firmer grounds as soon as the radio counts
for BL Lacs down to $\sim 50$ mJy will be available from DXRBS (Perlman et
al. 1998).

Ghisellini (1998), following the BeppoSAX detection of synchrotron peaks
at hard X-ray energies in the TeV BL Lacs MKN 501 (Pian et al. 1998) 
and 1ES2344+514 (Giommi et al. 1998), presented a scenario where the peak 
of the synchrotron emission in BL Lacs might reach very high energies, 
possibly well into the MeV region. Some of these objects, except perhaps 
the most extreme ones (which would be below our radio threshold) could 
be included in our sample, which is selected to have very high \fxfr ratios. 

Forty-eight objects in our sample have $f_{\rm x} > 10^{-11}$
\ergs~(thirty-five of these are known BL Lacs). Under the assumption
that the inverse Compton and synchrotron powers are of the same order
(see, e.g., Fossati et al. 1998), Stecker, de Jager \& Salamon (1996)
have shown that $\nu_{\rm TeV} f_{\rm TeV}
\sim \nu_{\rm x} f_{\rm x}$. These sources (at least the low-redshift one, 
because of intergalactic absorption) could then be detected at TeV frequencies
(especially during flares) even with the present generation of Cerenkov
telescopes. 

Finally, we stress that ``statistical'' methods of source identification of
the kind described in this paper (i.e., selecting rare sources with [almost]
unique spectral energy distribution or location in a multi-parameter,
multi-frequency space) will have to become more and more common in the near
future. Given the large number of up-coming surveys at various frequencies, in
fact, it will be increasingly difficult to conduct surveys the ``classical''
way, as the number and faintness of the objects involved will require
exceedingly large amounts of telescope time to secure optical spectroscopy and
identification. 

\section*{Acknowledgments}

P.G. acknowledges the STScI Visitor Program. This research has made use of the
following on-line services: the ASI-BeppoSAX SDC database and archive system;
the STScI Digitized Sky Survey; the ROE/NRL COSMOS service; the RGO/APM 
service; and the
NASA/IPAC National Extragalactic Database (NED). The Digitized Sky Surveys
were produced at the Space Telescope Science Institute under U.S. Government
grant NAG W-2166. The images of these surveys are based on photographic data
obtained using the Oschin Schmidt Telescope on Palomar Mountain and the UK
Schmidt Telescope. The plates were processed into the present compressed
digital form with the permission of these institutions. The UK Schmidt
Telescope was operated by the Royal Observatory Edinburgh, with funding from
the UK Science and Engineering Research Council (later the UK Particle Physics
and Astronomy Research Council), until 1988 June, and thereafter by the
Anglo-Australian Observatory. The blue plates of the southern Sky Atlas and
its Equatorial Extension (together known as the SERC-J), as well as the
Equatorial Red (ER), and the Second Epoch [red] Survey (SES) were all taken
with the UK Schmidt.

\label{lastpage}

\end{document}